\documentclass[11pt,a4paper,final]{iopart}
\usepackage{iopams}  
\usepackage{amssymb}  
\usepackage{graphicx}
\usepackage[breaklinks=true,colorlinks=true,linkcolor=blue,urlcolor=blue,citecolor=blue]{hyperref}
\begin{document}

\title[]{Wei-Norman-Kolokolov approach for Landau-Zener problems}

\author{M B Kenmoe$^{1}$}
\address{$^1$Mesoscopic and Multilayer Structures Laboratory, Faculty of Science, Department of Physics, University of Dschang, Cameroon}
\ead{kenmax15@yahoo.fr}

\author[cor1]{L C Fai$^{1}$}
\address{$^1$Mesoscopic and Multilayer Structures Laboratory, Faculty of Science, Department of Physics, University of Dschang, Cameroon}
\eads{corneliusfai@ictp.it}

\begin{abstract}
The tunneling between the $2S+1$  Zeeman multiplets of an arbitrary spin  $S$
 using the $SU(2)$ Lie group exponential ordering Wei-Norman technique is investigated. A spin subjected to a regular time-dependent magnetic field and a transverse colored noise is considered and the regimes of fast and slow noise examined. 
\end{abstract}
\vspace{0.5pc}

\submitto{\JPA}
\section{Introduction}
Following the rapid growth of the fields of spintronics and molecular electronics\cite{Bogani, Shirani, Loss}, there have been great achievements in modern and condensed matter physics. These new and exciting fields of physics whose interests for are manifold, help in explaining the mechanism of spin flip transitions and spin propagations\cite{Bett} 
in single- and multi- molecule magnets\cite{Sessoli1999, Sessoli2006, Garcia1997, sessoli1999.1}, quantum computation\cite{Loss1, Hanson} and
 Landau-Zener(LZ)\cite{lan, zen, stu, Majorana} transitions. The possibility of spin manipulation by means of an applied bias voltage has revealed  electron spin to be
  an information carrier in both classical and quantum information technology\cite{Miller, Brandford, Karimov}. A spin-$1/2$ particle is an appropriate two-level 
  quantum system.  This forms a quantum bit (qubit) being the basic unit of quantum information\cite{Thomas, Gradbert, Suter}. 
 
The full understanding of spin evolution in two-level atoms is necessary to establish predictions in $\mathcal{N}$-level atoms where more complicated patterns are expected as the total spin is $S$\cite{Kiselev}. A breakdown between 
theoretical and experimental expectations occurs when the spin dynamics become cumbersome\cite{Kiselev, Hao, Warren}. This is precisely a genuine signature of the complexity 
 of spin dynamics in synthesized nanomagnets\cite{Gross}. The fundamental Schr\"odinger equation for such situations involves $\mathcal{N}$ coupled linear 
  differential equations with nontrivial solutions. This difficulty is overcome by an alternative and rigorous  approach introduced by Wei and Norman (WN)\cite{wei, Dat1986, Dat1988} and corresponds to one of
 the Lie exponential forms. The WN approach involves the arrangement
  of the time-evolution operator (TEO) into a parametrized form that depends on the group dimension $n$. The time-dependent parametric functions 
 for $SU(2)$ symmetry group fully determined the TEO and are universal for all irreducible representations of the group\cite{Lie}. 
   
 Group theoretical requirements of $SU(n)$ are useful for quantum optics\cite{Sanchez, Elgin}, evolution of squeezed states\cite{Gerry}, propagation of Gaussian beams\cite{Dat1987}. 
 The present paper reveals the role of $SU(2)$ group in the physics of $\mathcal{N}$-energy levels evolving in time with the same symmetry and crossing at a single 
  point by linear variation of a controlling parameter (time, coordinate, energy, chemical potential, flux etc). The representation of the TEO that we consider was first proposed by Kolokolov\cite{kol} for functional integration of quantum magnets. 
  The application of additional transformations  permits to achieve the so-called Wei-Norman-Kolokolov (WNK) representation\cite{kol, Yuval}. 
  The case where the linear varying parameter is a regular field refers to as the LZ theory\cite{kay, kay1985, Schirmer, Saito2007}.

 The paper is presented as follows: Section \ref{Sec1} presents the TEO, its matrix elements (scattering matrix) properties; Section \ref{Sec2} is devoted to application to
  multi-level LZ transitions. Section \ref{Sec3} illustrates 
  the method for the case when an arbitrary spin $S$ is subjected to a regular magnetic field and a classical transverse noise with Gaussian realizations. 
	Here, the two limits of the noise being fast  or slow comparing the characteristic time scale of the noise with that of the system are considered.
	  Section \ref{Sec4} is the summary.
    
\section{Wei-Norman method for the $SU(2)$ finite dimensional Lie group}\label{Sec1}

\subsection{The time-evolution operator}\label{Sec1.1} 

Let $|\psi(t)\rangle$ be the Schr\"odinger state vector of an arbitrary spin $S$ at time $t$. The time evolution of the  coherently  driven system is described by the  propagator $\hat{U}(t,t_{0})=|\psi(t)\rangle\langle\psi(t_{0})|$, $(t>t_{0})$  which satisfies the time-dependent Schr\"odinger equation ($\hbar=1$):
 \begin{eqnarray}
i\frac{d}{dt}\hat{U}(t,t_{0})=\hat{\mathcal{H}}(t)\hat{U}(t,t_{0}),\label{equ1}
\end{eqnarray}
and the initial condition $\hat{U}(t_{0},t_{0})=\hat{\mathbf{1}}$ (where $\hat{\mathbf{1}}$ is a unit matrix).
 The Hamiltonian,
 \begin{eqnarray}\label{equ3}
\hat{\mathcal{H}}(t)=\sum_{j=1}^{3}\Theta_{j}(t)S_{j},
\end{eqnarray}
is  Hermitian and for arbitrary $t$ belongs to the finite dimensional Lie group $SU(2)$. 
Here, $\Theta_{j}(t)$ is a set of linearly independent complex functions of time.
The basis for the Lie space is chosen such that for all generators $S_{j}$ of $SU(2)$ we have $[S_{i},S_{j}]=c_{ij}^{\ell}S_{\ell}$ where $c_{ij}^{\ell}$ are constants structure on $SU(2)$. 

For our case, the fundamental solution $\hat{U}(t, t_{0})=\hat{\mathcal{T}}\exp[-i\int_{t_{0}}^{t}\hat{\mathcal{H}}(\tau)d\tau]$ (with $\hat{\mathcal{T}}$ being a time ordering operator) of equation (\ref{equ1}) belonging to the group $SO(3)$ is transferred to the TEO $\hat{U}(t)$   selected as the following Lie exponential form\cite{wei, kol, Yuval}:
\begin{eqnarray}
\hat{U}(t)=\exp({f(t)S_{-}})\exp({h(t)S_{z}})\exp({g(t)S_{+}}),\label{equ4}
\end{eqnarray}
where $S_{\pm}=S_{x}\pm iS_{y}$. Substituting the solution in equation (\ref{equ4}) into equation (\ref{equ1}), the functions $f(t)$, $h(t)$ and $g(t)$ 
are found to satisfy the following system of differential equations\cite{Yuval}:
\begin{eqnarray}\label{equ5}
\left\{
  \begin{array}{lll}
    i\dot{f}=\frac{1}{2}\Theta_{+}(t)-\Theta_{z}(t)f-\frac{1}{2}\Theta_{-}(t)f^{2},\\
    i\dot{h}=\Theta_{z}(t)+\Theta_{-}(t)f,\\
    i\dot{g}=\frac{1}{2}\Theta_{-}(t)e^{-h},
  \end{array}
\right.
\end{eqnarray}
assuming the initial values $f(t_{0})=h(t_{0})=g(t_{0})=0$. Here, $\Theta_{\pm}(t)=\Theta_{x}(t)\pm i\Theta_{y}(t)$; the dot on the functions refer to time derivatives. 
 Solutions of equation (\ref{equ5}) depend on the Riccati equation\cite{ric} (first equation in (\ref{equ5})) that can be transformed in to a second order differential equation\cite{ric}.

\subsection{Scattering matrix and properties}\label{Sec1.2} 
The state vector $|\psi(t)\rangle$ of an arbitrary spin $S$ describing $2S+1$ accessible or non accessible states in a
 time-dependent field may be decomposed as a direct sum
\begin{eqnarray}\label{equ6}
|\psi(t)\rangle=\sum_{m=-S}^{S}C_{m}(t)| S,m\rangle.
\end{eqnarray}
Here, $C_{m}(t)$ are amplitudes for finding a projection $m$ of $S$ on a given axis of the basis $| S, m\rangle$ ($m=-S, -S+1,...,S$).
This basis can be constructed with the aid of multispinor technique\cite{ Lie,QM}. We find the  matrix elements (scattering matrix):
\begin{eqnarray}\label{equ7}
U_{m,m'}^{S}(t, t_{0})= \langle S, m' |\hat{U}(t, t_{0})| S,m\rangle,
\end{eqnarray}
of the TEO $\hat{U}(t, t_{0})$. They relate the amplitudes $C_{m}(t)$ through the relation:
\begin{eqnarray}\label{equ8}
C_{m}(t)=\sum_{m'=-S}^{S}U_{m,m'}^{S}(t, t_{0})C_{m'}(t_{0}).
\end{eqnarray}
We choose the basis to make the states  mutually orthogonal ($\langle S, m| S, m'\rangle=\delta_{mm'} $) and the
closure relation $\sum_{m=-S}^{S}| S, m\rangle\langle S, m|=1$ always satisfies.
Then, assuming that the system evolves from $t_{0}=-\infty$, after some elaborated algebra, the matrix elements of the TEO defined in equation (\ref{equ4}) are obtained for arbitrary times $t$ as follows:
\begin{eqnarray}\label{equ9}
U_{m,m'}^{S}(Q)=\sqrt{\frac{(S-m)!(S+m')!}{(S+m)!(S-m')!}}\frac{g^{m'-m}e^{m'h}}{(m'-m)!}w^{S}_{mm'}(Q), \quad m'\geq m,
\end{eqnarray}
\begin{eqnarray}\label{equ10}
U_{m',m}^{S}(Q)=\sqrt{\frac{(S-m')!(S+m)!}{(S+m')!(S-m)!}}\frac{f^{m-m'}e^{m h}}{(m-m')!}w^{S}_{m'm}(Q), \quad m'\leq m,
\end{eqnarray}
and satisfy the closure relation
$\sum_{m'=-S}^{S}| U_{m,m'}^{S}(Q)|^{2}=1$ for fixed $m$;  $Q=(f, h, g)$. Here,
$w^{S}_{m,m'}(Q)={\empty}_{2}F_{1}(-S+m', S+m'+1, m'-m+1; -fge^{h})$ with ${\empty}_{2}F_{1}(...)$ 
being the Gauss hypergeometric function\cite{Erdelyi}. Also, if $f$, $h$ and $g$ are real, $[U_{m,m'}^{S}(f, h, g)]^{\dagger}=U_{m',m}^{S}(-g, -h, -f)$. 
If one assumes in this case that $Q^{\dagger}=(-g, -h, -f)$,  then $[U_{m,m'}^{S}(Q)]^{\dagger}=U_{m',m}^{S}(Q^{\dagger})$. Similarly, 
when $m=m'$, 
$U_{m,m}^{S}(Q)=e^{m h}w^{S}_{mm}(Q)$. Note that the transition amplitudes between the states with extremal spin projections are $U_{S,S}^{S}(Q)=e^{Sh}$ since $w^{S}_{SS}(Q)=1$ and $U_{-S,-S}^{S}(Q)=e^{-Sh}(1+e^{h}fg)^{2S}$. 
We present  $U_{m,m'}^{S}(Q)$ for some values of $S$: 
\begin{eqnarray}\label{equ10a}
\nonumber\hspace{2.5cm} U^{(1/2)}=
\left( {\begin{array}{*{20}c}
e^{h/2} &  ge^{h/2}\\
fe^{h/2} & e^{-h/2}(1+fge^{h})
\end{array} } \right),\\\hspace{0.7cm} 
U^{(1)}=
\left(
{\begin{array}{*{20}c}
e^{h} &  \sqrt{2}e^{h}g & e^{h}g^{2}\\
\sqrt{2}e^{h}f & 1+2e^{h}fg & \sqrt{2}g(1+e^{h}fg)\\
e^{h}f^{2} & \sqrt{2}f(1+e^{h}fg) & e^{-h}(1+fge^{h})^2
\end{array} } \right), \\\nonumber
\hspace{-2.5cm} U^{(3/2)}=\\\hspace{-2.8cm}\nonumber
\left(
{\begin{array}{*{20}c}
e^{3h/2} &        \sqrt{3}e^{3h/2}g       &     \sqrt{3}e^{3h/2}g^{2} &   e^{3h/2}g^{3}\\
\sqrt{3}e^{3h/2}f &   e^{h/2}(1+3fge^{h})       & e^{h/2}g(2+3fge^{h}) &   \sqrt{3}e^{h/2}g^{2}(1+fge^{h})\\
\sqrt{3}e^{3h/2}f^{2} &   e^{h/2}f(2+3fge^{h})      &  e^{-h/2}(1+fge^{h})(1+3fge^{h}) &  \sqrt{3}e^{-h/2}g(1+fge^{h})^{2}\\
e^{3h/2}f^{3} &  \sqrt{3}e^{h/2}f^{2}(1+fge^{h})  &  \sqrt{3}e^{-h/2}f(1+fge^{h})^2 &   e^{-3h/2}(1+fge^{h})^3 
\end{array} } \right).
\end{eqnarray}
The matrix element in position ($i, j$) corresponds to the transition amplitude from the  state with index $j$ to $i$.

With the help of the parametrization in equation (\ref{equ4}) and the Baker-Campbell-Haursdorff expansion for products of exponential operators, we compute
\begin{eqnarray}\label{equ11}
\hat{U}^{-1}(Q)\frac{\partial \hat{U}(Q)}{\partial Q}=\mathcal{A}(Q)\mathcal{G}(S).
\end{eqnarray}
Here,
\begin{eqnarray}\label{equ12}
\mathcal{A}(Q)=
\left(
{\begin{array}{*{20}c}
e^{h} &   -2ge^{h}       &    -g^{2}e^{h} \\
   0  &    1        &      g   \\
   0  &    0        &      1   \\
\end{array} } \right),
\end{eqnarray}
and $\mathcal{G}(S)=[S_{-}, S_{z}, S_{+}]^{T}$ is a vector whose elements are generators of the Lie algebra on $SU(2)$. This is useful to establish some properties of $\hat{U}(Q)$ and find the invariant integration measure\cite{Tung} associated with the 
parametrization (\ref{equ4}).  Thus, the invariant measure
$dQ=|\mathcal{A}(Q)|dfdhdg$  on the group $SU(2)$ reads
\begin{eqnarray}\label{equ13}
dQ=e^{h}dfdhdg.
\end{eqnarray}
This measure is normalized to the ''{\it group volume}'' $V_{group}=\int_{(\mathcal{C})} e^{h}dfdhdg$ such that $dQ=e^{h}dfdhdg/V_{group}$ is a normalized invariant measure. As a  consequence, the invariant measure $dQ$ permits the normalization of the matrix elements in equations (\ref{equ9}) and (\ref{equ10}) (see also reference \cite{Tung}):
\begin{eqnarray}\label{equ14}
\int dQ \Big[U^{j_{1}}_{m,k}(Q)\Big]\Big[U^{j_{2}}_{m', k'}(Q)\Big]^{-1}=\frac{1}{2j_{2}+1}\delta_{j_{1}j_{2}}\delta_{mm'}\delta_{kk'}.
\end{eqnarray} 
By expanding an arbitrary function of $Q$ into the basis of $U^{j}_{m, k}(Q)$ and making use of the integral relation (\ref{equ14}), it is demonstrated that:
\begin{eqnarray}\label{equ15a}
\sum_{jmk}(2j+1)\Big[U^{j}_{k, m}(Q)\Big]\Big[U^{j}_{m, k}(Q')\Big]^{-1}=\delta(Q-Q').
\end{eqnarray} 
Consider the positive $z$-axis. The system of equations (\ref{equ11}) contains three coupled equations for $\partial \hat{U}/\partial f$, $\partial \hat{U}/\partial h$ and $\partial \hat{U}/\partial g$. After decoupling, the functions $\hat{U}S_{-}$, $\hat{U}S_{+}$ and $\hat{U}S_{z}$ are extracted and sandwiched at the left and right hand sides by the state vector $|S,m\rangle$:    
 \begin{eqnarray}\label{equ15}
\hat{\mathcal{L}}_{z}U_{m,m'}^{S}(Q)=mU_{m,m'}^{S}(Q),
\end{eqnarray} 
\begin{eqnarray}\label{equ16}
\hat{\mathcal{L}}_{\pm}U_{m,m'}^{S}(Q)=\sqrt{(S\mp m)(S\pm m+1)} U_{m\pm1,m'}^{S}(Q),
\end{eqnarray}
where
\begin{eqnarray}\label{equ17}
\hat{\mathcal{L}}_{+}=\partial_{g},
\quad 
\hat{\mathcal{L}}_{z}=\partial_{h}-g\partial_{g},
\quad
\hat{\mathcal{L}}_{-}=e^{-h}\partial_{f}+2g\partial_{h}-g^{2}\partial_{g},
\end{eqnarray} 
and $\partial_{Q}=\partial/\partial Q$. For the negative $z$-axis, we have
 \begin{eqnarray}\label{equ18}
\check{\mathcal{L}}_{-}=\partial_{f},
\quad
\check{\mathcal{L}}_{z}=-\partial_{h}+f\partial_{f},
\quad
\check{\mathcal{L}}_{+}=e^{-h}\partial_{g}+2f\partial_{h}-f^{2}\partial_{f},
\end{eqnarray}
satisfying the following commutation relations
\begin{eqnarray}\label{equ19}
 [\mathcal{L}_{+},\mathcal{L}_{-}]=2\mathcal{L}_{z},\quad [\mathcal{L}_{z},\mathcal{L}_{\pm}]=\pm \mathcal{L}_{\pm},
\end{eqnarray} 
where $\mathcal{L}=(\hat{\mathcal{L}}, \check{\mathcal{L}})$ are isomorphic to the $SU(2)$ group according to (\ref{equ19}). 
On the other hand, $\check{\mathcal{L}}_{\pm}$ satisfy similar relations as $\hat{\mathcal{L}}_{\pm}$, except $\check{\mathcal{L}}_{z}$ that obeys:
\begin{eqnarray}\label{equ20}
\check{\mathcal{L}}_{z}U_{m,m'}^{S}(Q)=-m'U_{m,m'}^{S}(Q).
\end{eqnarray} 
One proves that the invariant Casimir operator
\begin{eqnarray}\label{equ21}
\hat{\mathcal{L}}^{2}=e^{-h}\partial_{fg}+\partial_{hh}+\partial_{h},
\end{eqnarray}
commutes with all operators in equations (\ref{equ17}) and (\ref{equ18}) and satisfies the eigenvalues equation
\begin{eqnarray}\label{equ22}
\hat{\mathcal{L}}^{2}U_{m,m'}^{S}(Q)=S(S+1)U_{m,m'}^{S}(Q).
\end{eqnarray}
By writing out the $f$-, $h$- and  $g$-dependence in equations  (\ref{equ9}) and (\ref{equ10}), we obtain the differential equations for 
 $w^{S}_{m,m'}(Q)$ when $m'\geq m$,
\begin{eqnarray}\nonumber\label{equ23}
\Big\{e^{-h}\partial_{f}\Big[\partial_{g}+[m'-m]\frac{1}{g}\Big]+\partial_{h}\Big[\partial_{h}+[2m'+1]\Big]\\\hspace{5cm}-(S-m')(S+m'+1)\Big\} w_{m,m'}^{S}(Q)=0,
\end{eqnarray}
and
\begin{eqnarray}\nonumber\label{equ24}
\Big\{e^{-h}\partial_{g}\Big[\partial_{f}+[m-m']\frac{1}{f}\Big]+\partial_{h}\Big[\partial_{h}+[2m+1]\Big]\\\hspace{5cm}-(S-m)(S+m+1)\Big\} w_{m',m}^{S}(Q)=0,
\end{eqnarray}
when $m'\leq m$. The change of variable $z=-fge^{h}$ yields the equation for Gauss hypergeometric function $_{2}F_{1}(...)$ respectively related to 
 equations  (\ref{equ23}) and (\ref{equ24}):
\begin{eqnarray}\nonumber\label{equ25}
\Big\{z(1-z)\frac{d^{2}}{d z^{2}}+\Big[(m'-m+1)-2(m'+1)z\Big]\frac{d}{d z}\\\hspace{5cm}+(S-m')(S+m'+1)\Big\}w_{m,m'}^{S}(z)=0,
\end{eqnarray}
\begin{eqnarray}\nonumber\label{equ26}
\Big\{z(1-z)\frac{d^{2}}{d z^{2}}+\Big[(m-m'+1)-2(m+1)z\Big]\frac{d}{d z}\\\hspace{5cm}+(S-m)(S+m+1)\Big\}w_{m',m}^{S}(z)=0.
\end{eqnarray}
Relations (\ref{equ21})-(\ref{equ26}) are valid for $\check{\mathcal{L}}$ as $\hat{\mathcal{L}}^{2}=\check{\mathcal{L}}^{2}$. Thus, equation (\ref{equ22}) leads to a unique equation which holds for all $m$ and $m'$:
\begin{eqnarray}\label{equ27}
\Big\{z(1-z)\frac{d^{2}}{d z^{2}}+\Big[1-2z\Big]\frac{d}{d z}-S(S+1)\Big\}U_{m,m'}^{S}(z)=0.
\end{eqnarray}
An alternative form of  equation (\ref{equ22}) is obtained by changing 
 $U_{m,m'}^{S}(Q)\to e^{-h}\tilde{U}_{m,m'}^{S}(Q)$. This procedure leads us to,
 \begin{eqnarray}\label{equ28}
\Big(e^{-h}\partial_{fg}+\partial_{hh}-\partial_{h}-S(S+1)\Big)\tilde{U}_{m,m'}^{S}(Q)=0.
\end{eqnarray} 
The properties (\ref{equ9})-(\ref{equ28}) form a solid background to discuss problems of transitions
between the $2S+1$ multiplets of a spin $S$ placed in a time-dependent magnetic field. This reminds us of the universal Wigner
 $D$-matrices\cite{Wigner}.The algebra of these matrices span into the $SU(2)$ group and is frequently used to describe spin dynamics.
A similar system of three equations like (\ref{equ5}) associated with the parametrization of the
TEO (rotation operator) with time-dependent Euler's angles involves
complicated combinations of trigonometric functions with non trivial solutions\cite{Hogan}.
The contrast with the parametrization (\ref{equ4}) becomes evident as the system of equation (\ref{equ5}) depends on a known equation (Riccati equation) with computable solutions.  Some examples are shown below for illustration.

\section{Landau-Zener transitions for an arbitrary spin $S$}\label{Sec2}

Transitions between energy levels at avoided crossings are known as LZ transitions. 
We write down the TEO for the LZ model generalized to the $2S+1$ levels of a spin $S$. 
The vector $\vec{\Theta}(t)$ is identified by
$\vec{\Theta}_{\rm LZ}(t)=(2\Delta, 0, 2 v t)$ where $v$ is the Zeeman constant sweep velocity, $\Delta$ the coupling between split levels. 

For the sake of  generalizations of statements and from the $2S+1$ dimensional representation of the universal enveloping algebra of
$SU(2)$, the model is written in the basis of the irreducible spin tensors $\hat{T}_{LM}^{(S)}$ as follows:
\begin{eqnarray}\label{equ29}
\hat{\mathcal{H}}_{\rm LZ}(t)=\sum_{L=0}^{2S}\sum_{M=-L}^L\alpha_{L,M}^{(S)}(t)\hat{T}_{LM}^{(S)}.
\end{eqnarray}
$\alpha_{L,M}^{(S)}(t)$ are linearly independent complex functions of time. The matrix elements of $\hat{T}_{LM}^{(S)}$ in the basis $| S,m \rangle$ are calculated with the aid of the Wigner-Eckart\cite{QM, DA} theorem and are written as follows:
 \begin{eqnarray}\label{equ30}
\Big[\hat{T}_{LM}^{(S)}\Big]_{mm'}=\sqrt{\frac{2L+1}{2S+1}}C_{S,m',L, M}^{S, m},
\end{eqnarray}
where $C_{S,m',L, M}^{S, m}=\langle Sm',LM | | S,m \rangle$ are Clebsch-Gordan coefficients\cite{QM, DA}.
The linearity of the LZ Hamiltonian and the Wigner-Eckart\cite{QM, DA} theorem impose $L=0$ and $L=1$ in equation (\ref{equ29}). Thus, 
\begin{eqnarray}\label{equ31ab}
\nonumber\alpha_{0,0}^{(S)}=0, \quad \alpha_{1,-1}^{(S)}=\sqrt{2}\Delta A^{(S)}, \quad \alpha_{1,+1}^{(S)}=-\sqrt{2}\Delta A^{(S)},\quad \alpha_{1,0}^{(S)}=2\alpha tA^{(S)},\\
\end{eqnarray}
where
\begin{eqnarray}\label{equ31ac}
A^{(S)}=\sqrt{\frac{S(S+1)(2S+1)}{3}}.
\end{eqnarray}

The TEO is fully determined once the Riccati equation is solved. We transform the Riccati equation in  (\ref{equ5}) to a second order differential equation with the transformation $f(t)=\frac{i}{\Delta}\frac{d}{dt}\ln Z(t)$  where
$Z(t)=\mathcal{Y}(\xi)\exp\Big(\frac{i}{2}\int^{t}_{t_{0}}\Theta_{z}(\tau)d\tau\Big)$. The function $\mathcal{Y}(\xi)$ satisfies the parabolic cylinder (Weber's) equation\cite{Kiselev, Erdelyi, Daniel}:
\begin{eqnarray}\label{equ32}
\frac{d^{2}}{d\xi^{2}}\mathcal{Y}(\xi)+\Big[i\delta-\frac{1}{2}-\frac{\xi^{2}}{4}\Big]\mathcal{Y}(\xi)=0.
\end{eqnarray}
Here,  $\xi=\sqrt{2v}t e^{-i\pi/4}$ and $\delta=\Delta^{2}/v$.
We finally obtain $f(t)$, $h(t)$ and $g(t)$ in terms of $\mathcal{Y}(\xi)\equiv \mathcal{Y}(t)$ as follows\cite{Kiselev}:
\begin{eqnarray}\label{equ33}
f(t)=\frac{i}{\Delta}\frac{d}{dt}\ln \mathcal{Y}(t)-\frac{v t}{\Delta}, \quad h(t)=2\ln \mathcal{Y}(t),
\quad g(t)=-i\Delta\int_{t_{0}}^{t}\frac{dt'}{\mathcal{Y}^{2}(t')}.
\end{eqnarray}
The parameters $f(t)$, $h(t)$ and  $g(t)$ entirely determine the TEO.  They do not contain
 any representation index of the group $SU(2)$. They are universal for all irreducible representation of $SU(2)$. This is a generalization 
 of Majorana\cite{Majorana} solutions (solutions in terms of parabolic cylinder functions). LZ transitions probabilities depend only on $S$ and $\delta$.
 
Equation (\ref{equ32}) has four solutions\cite{Erdelyi},  with an appropriate choice of the solution which satisfies $\mathcal{Y}(-\infty)=1$,
\begin{eqnarray}\label{equ34}
f(\infty)=\chi e^{-i\Phi},\quad h(\infty)=-\pi\delta, \quad g(\infty)=-\chi e^{i\Phi},
\end{eqnarray}
where $\chi=[2\sinh(\pi\delta/2)]^{1/2}e^{\pi\delta/4}$ and $\Phi=\frac{\pi}{4}+\arg\Gamma(1-i\delta)+\delta(\ln\delta-1)$
is the Stoke's phase originating from large expansion of parabolic cylinder functions\cite{Crothers, kay1997}. 

On the other hand, the product of two Gauss hypergeometric functions leads to Appell's function $F_{4}[...]$ \cite{Watson, Wimp}. 
This permits to arrive the generalized formula:
\begin{eqnarray}\label{equ36}
\mathcal{P}^{(S)}_{m'\to m}(\infty)=\frac{(S+m')!}{(S-m')!} \frac{a^{m'-m}b^{m'+m}}{(m'-m)!(m'+m)!}\ell^{S}_{mm'},
\end{eqnarray}
for $m'\geq m$ where 
\begin{eqnarray}\label{equ37}
\ell^{S}_{m,m'}=F_{4}(-S+m', S+m'+1, m'-m+1, m'+m+1; a^{2}; b^{2}),
\end{eqnarray}
$a=1-e^{-\pi\delta}$ and $b=e^{-\pi\delta}$. We achieve our goal. The case  $m'\leq m$ is obtained from (\ref{equ36}) by swapping $m\leftrightarrows m'$ everywhere they appear.
 This compact formula may be used to work out the transition probabilities for exceedingly large values of $S$. 

We present some limiting cases. For non-adiabatic transitions between the states with extremal spin projections, $\mathcal{P}^{(S)}_{S\to S}(\infty)=b^{2S}$ and $\mathcal{P}^{(S)}_{S\to -S}(\infty)=a^{2S}$. We achieve the LZ formula when $S=1/2$. Similarly, for non-adiabatic transitions between the state with maximum spin projection and other projections, $\ell^{S}_{m,S}=1$. The transition probabilities are given by
\begin{eqnarray}\label{equ37a}
\mathcal{P}^{(S)}_{S\to m}(\infty)=\frac{a^{S-m}b^{S+m}}{(S-m)!(S+m)!}(2S)!.
\end{eqnarray}
These probabilities are independent and obey the condition $\sum_{m=-S}^{S}\mathcal{P}^{(S)}_{S\to m}(\infty)=1$. The probabilities for transitions between intermediate states can be calculated from (\ref{equ36}).

\section{Transverse noise on an arbitrary spin $S$}\label{Sec3}

The spin $S$ interacts now with a noisy environment (host lattice;  nuclear spins that create a field of random amplitude and direction, the Overhauser field associated with a classical noise with Gaussian realizations). The transverse components of the magnetic field
randomly fluctuate while its longitudinal component oscillates with a frequency $\omega(t)=2 v t$:
\begin{eqnarray}\label{equ38}
\vec{\Theta}(t)=(\mathcal{J}_{x}(t), \mathcal{J}_{y}(t), \omega(t)).
\end{eqnarray} 
The fields $\mathcal{J}_{i}(t)$ ($x,y$) are realized as zero-mean Gaussian noise with two-time correlation functions given by 
$\langle \mathcal{J}_{i}(t)\mathcal{J}_{j}(t')\rangle=\mathcal{R}_{ij}(\gamma| t-t'|)$ where
\begin{eqnarray}\label{equ39}
\mathcal{R}_{ij}(\gamma| t-t'|)=\eta^{2}\exp(-\gamma|t-t'|)\delta_{ij}.
\end{eqnarray}
 Here, $\eta$ is the noise strength while $\gamma$ is the inverse characteristic correlation time of the noise. We discuss noise-induced transitions in the frame 
 of equation (\ref{equ5}) and  construct the relevant Fokker-Planck (FP) equation \cite{Risken, Gardiner, kalm}. 
 Equation (\ref{equ5}) however contains fast oscillating terms which should be removed. 
  To do that, we introduce slowly oscillating variables,
 \begin{eqnarray}\label{equ40}
 f(t)\to \tilde{f}(t)\exp\Big(i\int^{t}_{t_{0}}\Theta_{z}(\tau)d\tau\Big),  h(t)\to \tilde{h}(t)-i\int^{t}_{t_{0}}\Theta_{z}(\tau)d\tau,  g(t)\to \tilde{g}(t).\nonumber\\
 \end{eqnarray}
 The stochastic Langevin variables become $Q(t)=(\tilde{f}(t), \tilde{h}(t), \tilde{g}(t))$.
  We examine in the next sections two limiting cases of noise.

\subsection{Fast noise for arbitrary spin $S$}\label{Sec3.1}

For this case, $\gamma\to\infty$, the characteristic time scale of the system is exceedingly large compared
to that of the noise. With the help of Novikov's theorem\cite{Novikov}, a relevant FP equation is derived for the probability density function (pdf) $\mathbf{W}(Q, t)$ as follows(see \ref{App} for details): 
 \begin{eqnarray}\label{equ41}
\partial_{t}\mathbf{W}(Q, t)=-\Omega(t)\check{\mathcal{L}}_{FP}(Q) \mathbf{W}(Q,t),
\end{eqnarray}
with the initial condition $\mathbf{W}(Q, t_{0})=\delta(Q-Q_{0})$ where $Q_{0}$ is an initial configuration; $\Omega(t)$ is the noise spectral density capturing information about the environment (see \ref{App}).
\begin{eqnarray}\label{equ43}
\check{\mathcal{L}}_{FP}(Q)=e^{-h}\partial_{fg}+2\partial_{fh}f-\partial_{ff}f^{2}+\partial_{f}f-\partial_{h},
\end{eqnarray}
a FP-like operator. From equations (\ref{equ17})-(\ref{equ22}), equation (\ref{equ43}) may be transformed to the following:
\begin{eqnarray}\label{equ44}
\check{\mathcal{L}}_{FP}(Q)=\mathcal{L}_{FP}(Q)-(\check{\mathcal{L}}_{z}+1)^{2},
\end{eqnarray}
where 
\begin{eqnarray}\label{equ45}
\mathcal{L}_{FP}(Q)=e^{-h}\partial_{fg}+\partial_{hh}-\partial_{h}.
\end{eqnarray}
The operator $(\check{\mathcal{L}}_{z}+1)$ in equation (\ref{equ44}) enters the FP operator $\check{\mathcal{L}}_{FP}$ as a consequence of the non-stochastic character of the magnetic field along the direction of the sweeping field. For its random distribution it is found to vanish. Interestingly, $\mathbf{W}(Q, t)$ in the FP equation will be entirely related to $\tilde{U}^{L}_{M, M'}(Q)$. It is instructive to note that:
\begin{eqnarray}\label{equ46}
(\check{\mathcal{L}}_{z}+1)\tilde{U}^{L}_{M, M'}(Q)=-M'\tilde{U}^{L}_{M, M'}(Q),
\end{eqnarray}
whereby, $\check{\mathcal{L}}_{FP}(Q)$ has the eigenvalues $L(L+1)-M'^{2}$ and the eigenfunctions $e^{h} U^{L}_{M, M'}(Q)$.
 After solving equation (\ref{equ41}), one obtains $2S+1$ solutions. The final solution is expressed as a linear combination of all particular solutions with $2S+1$ constants. The correct solution should satisfy the initial condition $\mathbf{W}(Q, t_{0})=\delta(Q-Q_{0})$. The normalization constants  that enters the final solution are determined with the aid of property (\ref{equ15a}). The pdf is then  obtained as follows:
\begin{eqnarray}\label{equ47}
\mathbf{W}(Q, t)=\sum_{LMM'} (2L+1)e^{-[L(L+1)-M'^{2}]\theta(t)/4} \Big[U^{L}_{M', M}(Q_{0})\Big]^{-1}\Big[U^{L}_{M, M'}(Q)\Big],
\end{eqnarray}
where, $\theta(t)=4\int_{t_{0}}^{t}\Omega(t')dt'$ is associated with the phase accumulated in an interval $(t_{0}, t]$. 
For a sweeping varying from $t_{0}=-\infty$ to $t=\infty$, one finds that
 $\theta(\infty)=4\pi\mathcal{R}(0)/v$ (see Reference \cite{Kenmoe} for more details). 

We compute the probability for an arbitrary spin $S$ initially in a configuration $Q_{0}$ at time $t_{0}$ to 
end up at a different configuration $Q$ at time $t$ by expanding the density matrix into 
the basis of the irreducible spin tensors $\hat{T}_{LM}^{(S)}$ (analogous to spherical harmonics) \cite{Blum}:
\begin{eqnarray}\label{equ48}
\hat{\rho}(t)=\sum_{L=0}^{2S}\sum_{M=-L}^L\langle\hat{T}_{LM}^{(S)}\rangle(t) \hat{T}_{LM}^{(S)}.
\end{eqnarray}
The tensors operators $\hat{T}_{LM}^{(S)}$ form an orthogonal basis in the $(2S+1)\times (2S+1)$ space of the density matrix which corresponds to $SO(3)$ rather than $SU(2)$ group. The expansion (\ref{equ48}) is valid for any Hermitian matrix with a trace equals to $1$. The brackets $\langle...\rangle$ express $\textmd{Tr}(...\rho)$, the coefficients  $\langle\hat{T}_{LM}^{(S)}\rangle(t)$ of expansion (\ref{equ48}) are Bloch tensors. They evolve independently for all given values of $M$ according to an equation 
which is analogous to the optical Bloch equation for a Bloch vector set on a unit sphere. $\langle\hat{T}_{LM}^{(S)}\rangle(t)$ can also be viewed as an 
average of $\hat{T}_{LM}^{(S)}(t)$ over environmental ($env$) degrees of freedom when $\hat{\rho}\to\hat{\rho}_{env}$(reduced density matrix). For our case, $\langle...\rangle$ 
accounts only for spin fluctuations, $\langle...\rangle\to\langle...\rangle_{phase}$ with
\begin{eqnarray}\label{equ49}
\langle...\rangle_{phase}(t)=\int dQ...\mathbf{W}(Q,t).
\end{eqnarray}

The distribution $\mathbf{W}(Q, t)$ is spatially uniform (no spatial decoherence). We find the Bloch tensors for $S$ in the initial configuration $Q=Q_{0}$. 
The procedure in equation (\ref{equ49}) yields
 \begin{eqnarray}\label{equ50}
\langle\hat{T}_{LM}^{(S)}\rangle(t)=\hat{T}_{LM}^{(S)}e^{-[L(L+1)-M^2]\theta(t)/4}. 
 \end{eqnarray}
The diagonal components of the Bloch tensors (\ref{equ50}) are non vanishing only for zero projection ($M=0$) of $S$ onto the direction of the sweeping field. The non diagonal components are different from zero if there is a coherence in the initial
state. For a different projection ($M\neq0$), the Bloch tensors  vanish as a result
of averaging over the random phases in the initial state. The diagonal components $\langle\hat{T}_{L0}^{(S)}\rangle(t)$ of the Bloch tensors involve 
$S_{z}$, $S_{z}S_{z}$,...They are combinations of diagonal components $\hat{\rho}_{jj}$ of the density matrix (transition probabilities) and represent population differences between levels. For instance, 
$\langle\hat{T}_{L0}^{(1/2)}\rangle(t)=\sqrt{2}( \rho_{11}(t)-\rho_{22}(t))$ is the population difference for two-level systems. 
 A key observation related to (\ref{equ50}) is that the diagonal components $\langle\hat{T}_{L0}^{(S)}\rangle(t)$ of the Bloch tensors  decrease monotonously with
time due to the noise.  This remark shows that the population differences
in average can only decrease after transitions. They may increase if there is a coherence in the initial
state as non diagonal components of the Bloch tensors are not zero.

Considering equation (\ref{equ15a}) and (\ref{equ48}) then, the transition probability is obtained as follows: 
 \begin{eqnarray}\label{equ52}
P_{\textmd{fn}}^{(S)}[m\to m']=\frac{1}{2S+1}+\sum_{L=1}^{2S}\sqrt{\frac{2L+1}{2S+1}}\Big[\hat{T}_{L0}^{(S)}\Big]_{mm} C_{S,m', L, 0}^{S, m'}
e^{-L(L+1)\theta/4}.
 \end{eqnarray}
 Similar results were obtained in Reference \cite{pok}. When $S=1/2$ and $S=1$ we reproduce the results in references \cite{Kenmoe, pok2003, pok}.
For illustration, Table\ref{TAB1} shows the transition  probabilities between Zeeman multiplets when $S=3/2$. The remaining 
 part of the table being completed using symmetry between levels or identical procedure.
Equation (\ref{equ52}) is obtained with the help of $\hat{T}_{00}^{(S)}$ \cite{Blum, pok}. 
\begin{table}{}
\centering
\begin{tabular}{lcr}
\hline
\hline
State-to-state &  Final occupation for $t=\infty$ \\
\hline
\hline

$|\frac{3}{2},\frac{3}{2}\rangle\to|\frac{3}{2},\frac{3}{2}\rangle$ & $\frac{1}{4}\Big[1+\frac{1}{5}e^{-3\theta}+e^{-3\theta/2}+\frac{9}{5}e^{-\theta/2}\Big]$ \vspace{0.1cm}\\ 
$|\frac{3}{2},\frac{3}{2}\rangle\to|\frac{3}{2},\frac{1}{2}\rangle$ & $\frac{1}{4}\Big[1-\frac{3}{5}e^{-3\theta}-e^{-3\theta/2}+\frac{3}{5}e^{-\theta/2}\Big]$\vspace{0.1cm}\\ 
$|\frac{3}{2},\frac{3}{2}\rangle\to|\frac{3}{2},-\frac{1}{2}\rangle$ & $\frac{1}{4}\Big[1+\frac{3}{5}e^{-3\theta}-e^{-3\theta/2}-\frac{3}{5}e^{-\theta/2}\Big]$\vspace{0.1cm}\\
$|\frac{3}{2},\frac{3}{2}\rangle\to|\frac{3}{2},-\frac{3}{2}\rangle$ & $\frac{1}{4}\Big[1-\frac{1}{5}e^{-3\theta}+e^{-3\theta/2}-\frac{9}{5}e^{-\theta/2}\Big]$
\\ \\
$|\frac{3}{2}, -\frac{1}{2}\rangle\to|\frac{3}{2},\frac{3}{2}\rangle$ & $\frac{1}{4}\Big[1+\frac{3}{5}e^{-3\theta}-e^{-3\theta/2}-\frac{3}{5}e^{-\theta/2}\Big]$ \vspace{0.1cm}\\ 
$|\frac{3}{2}, -\frac{1}{2}\rangle\to|\frac{3}{2},\frac{1}{2}\rangle$  & $\frac{1}{4}\Big[1-\frac{9}{5}e^{-3\theta}+e^{-3\theta/2}-\frac{1}{5}e^{-\theta/2}\Big]$\vspace{0.1cm}\\
$|\frac{3}{2}, -\frac{1}{2}\rangle\to|\frac{3}{2},-\frac{1}{2}\rangle$ & $\frac{1}{4}\Big[1+\frac{9}{5}e^{-3\theta}+e^{-3\theta/2}+\frac{1}{5}e^{-\theta/2}\Big]$\vspace{0.1cm}\\
$|\frac{3}{2}, -\frac{1}{2}\rangle\to|\frac{3}{2},-\frac{3}{2}\rangle$  & $\frac{1}{4}\Big[1-\frac{3}{5}e^{-3\theta}-e^{-3\theta/2}+\frac{3}{5}e^{-\theta/2}\Big]$ \vspace{0.1cm}\\
\hline
\hline
\end{tabular}
\caption{{\small Landau-Zener transition probabilities for some of the four multiplets of the spin $S=3/2$. (see also references \cite{pok2003,pok})}}\label{TAB1}
\end{table}

\subsection{Slow noise for arbitrary spin S}\label{Sec3.2}
 
For the case of slow noise, $\gamma \to 0$, the correlation time is exceedingly greater than the characteristic time of the LZ process. A FP equation is {\it not} easily achieved. We readily solve the Langevin equations in phase space ($f, h, g$). We consider the long-time asymptotic transition probabilities when $t\to\infty$
 and perform Gaussian ensemble average $\langle...\rangle_{noise}$ (given in reference \cite{Kenmoe}) over the distribution
$\mathcal{Q}$ of the noise\cite{Kenmoe} i.e.:
 \begin{eqnarray}\label{equ54}
P_{sn, \beta}^{(S)}[m'\to m]=\langle \mathcal{P}^{(S)}_{m'\to m}(\mathcal{Q})\rangle_{noise}.
\end{eqnarray}
For a single component transverse noise ($X$-noise, $\beta=X$), $\mathcal{Q}=x$ while for two-component ($XY$-noise, $\beta=XY$),  $\mathcal{Q}=\sqrt{x^2+y^2}$ and the average 
$\langle...\rangle_{noise}$ is transferred to a double Gaussian distribution (see reference \cite{Kenmoe}). For transitions between the state  with maximum spin projection $(m'=S)$ and that with other projections $(m\le -S)$, the procedure (\ref{equ54})  accounting for (\ref{equ37a}) leads to
\begin{eqnarray}\label{equ54c}
P_{sn, \beta}^{(S)}[S\to m]=\sum_{L=0}^{S-m}\frac{(-1)^{L}(2S)!}{L!(S+m)!(S-m-L)!}\Big[1+\frac{2\pi \eta^{2}}{v}(S+m+L)\Big]^{-\frac{\kappa}{2}}.\hspace{0.7cm}
\end{eqnarray}
Here, $\kappa$ takes two values $1$ and $2$ for single- and two-component  transverse noise respectively.
Some results illustrating this procedure are presented below for $S=1/2$, $S=1$ and $S=3/2$:
\begin{eqnarray}\label{equ55}
P_{sn}^{(1/2)}=
\left( {\begin{array}{*{20}c}
p &  q\\
q & p
\end{array} } \right),\\
P_{sn}^{(1)}=
\left(
{\begin{array}{*{20}c}
 p^{2}     &       2pq      &     q^{2} \\
   2pq    &      (2p-1)^{2}    &       2pq\\
  q^{2}   &       2pq      &        p^{2} 
\end{array} } \right),\\
P_{sn}^{(3/2)}=
\left( {\begin{array}{*{20}c}
   p^{3}     &       3p^{2}q  &    3pq^{2}    &   q^{3} \\
3p^{2}q   &      (3p-2)^{2}p   &  (3p-1)^{2}q  &  3pq^{2} \\
3pq^{2}   &   (3p-1)^{2}q  &     (3p-2)^{2}p       &  3p^{2}q\\
  q^{3}   &       3pq^{2}  &    3p^{2}q    &     p^{3} 
\end{array} } \right).
\end{eqnarray}
Here, $q=1-p$ with $p^{\nu}=[1+\frac{2\pi\nu\eta^{2}}{v}]^{-\frac{\kappa}{2}}$. It is instructive to note that when using $q$ and $p$ we first collect all powers of $p$. For instance $pq=p-p^{2}$ or $pq^{2}=p-2p^{2}+p^{3}$.

Slow noise transforms the Gaussian properties of  LZ transitions probabilities to Lorentzian\cite{Kenmoe}.  

\section{Summary}\label{Sec4}

 We have developed an approach to problems of transitions in $\mathcal{N}$-level systems induced by linearly varying protocols (time, coordinate, energy, chemical potential, flux etc) and(or) fluctuations of Overhauser field associated with hyperfine interactions. The approach is applied to cases of transitions induced by random transverse (or regular) magnetic fields with focus on level-crossing systems
 with $SU(2)$  symmetry. The universal model Hamiltonian that considers all irreducible representations of an arbitrary spin $S$  is achieved by expressing the Landau-Zener 
 Hamiltonian through spin tensors which envelop the universal $su(2)$ algebra.

The fundamental solution to the time-dependent Schr\"odinger equation is expanded as a product of three exponential operators (with three time-dependent functions $f(t)$, $h(t)$ and $g(t)$). We indicate the condition under which the TEO is fully determined, namely once the function $f(t)$  (solution of a Riccati equation) is obtained. For general $SU(2)$ Landau-Zener problem, this equation is analytically solved, $f(t)$, $h(t)$ and $g(t)$ are expressed  through Weber's parabolic cylinder  functions. The finger prints of $SU(2)$ irreducible representation indices disappering. This is observed to be a generalization  of Majorana solutions\cite{Majorana} for two-level crossing. A compact formula expressing LZ transitions between diabatic states is derived through the Appell function $F_{4}(...)$\cite{Watson, Wimp}. The expressions of the $SU(2)$ "scattering
matrices" and their useful properties are presented.

Considering noise-induced Landau-Zener transitions, the set of 
equations for $f(t)$, $h(t)$ and $g(t)$ are observed to be described by nonlinear Langevin equations. In the limit of fast noise, a relevant Fokker-Planck equation for the probability density function $\mathbf{W}(Q,t)$ is derived. The properties of the scattering matrix elements $U^{S}_{m,m'}(t)$ help to solve this equation exactly in phase space $(f,h,g)$ where spin dynamics are described. The originality of the approach presented in this paper consists in its  generalization. Rather than solving the von-Neuman equation for the density matrix we have expanded the latter in the basis of irreducible spin tensors. The coefficients of this expansion (Bloch tensors) were obtained  by analytically solving a Fokker-Planck
equation. In the expression for probabilities of diabatic transitions between  
 Zeeman multiplets we introduce functions with clear physical meaning such as Bloch tensors. We express our results in terms of well-know mathematical functions such as the Appell function $F_{4}(...)$\cite{Watson, Wimp}. It is observed that all transition probabilities are exclusively  tailored by $S$  and(or) $\Delta^2/v$($\eta^2/v$). The results for fast noise are observed to be in  excellent agreement with those in reference\cite{pok2003, pok}. In the opposite limit of slow noise, the nonlinear Langevin equations are solved and a  Bloch average procedure\cite{Kenmoe} is adopted.

The present study helps in explaining the mechanism of spin-flip transitions and spin-propagations\cite{Bett} in single- and multi-molecule magnets, synthesis and magnetic measurements of cubic nanomagnets\cite{Sessoli1999, Sessoli2006, Garcia1997, sessoli1999.1}, quantum computation\cite{Loss1, Hanson} and Landau-Zener transitions \cite{lan, zen, stu, Majorana}. The study might also be extended to different types of magnetic fields (not necessary linearly varying) and lead to solving a nonlinear Ricatti equation\cite{ric}.

\appendix
\section{Fokker-Planck equation}\label{App}
We write a Fokker-Planck (FP) equation for an arbitrary Gaussian process in the limit of fast noise. We use Einstein's notation. We consider Langevin\cite{Risken} equations derived from TEOs for the group $SU(2)$. Three generators are involved and result in a three-dimensional Langevin equation of the form\cite{Risken}
\begin{eqnarray}\label{eqA1}
\dot{Q_{\mu}}(t)=A_{\mu}(Q,t)+B_{\mu k}(Q,t)\mathcal{J}_{k}(t),
\end{eqnarray}
$\mu=1,2,3$. For the case when the particle is placed in a  magnetic field random in the transverse directions and regular in the longitudinal direction with a frequency $\Theta_{z}(t)$, the equation (\ref{eqA1}) contains fast oscillating terms that are eliminated by defining an appropriate gauge transformation.   The gauge in such cases, consists of tracing out $\Theta_{z}(t)$ in $A_{\mu}(Q,t)$ (see equation (\ref{equ40})). After doing this,  we further consider only slow variables $\tilde{Q}(t)$ and omit the sign tilde on $Q(t)$. The noise fields $\mathcal{J}_{x,y}(t)$ are transformed as 
\begin{eqnarray}\label{eqA2}
\tilde{\mathcal{J}}_{x}(t)=\mathcal{J}_{x}(t)\cos\vartheta(t)+\mathcal{J}_{y}(t)\sin\vartheta(t),
 \end{eqnarray}
\begin{eqnarray}\label{eqA3}
\tilde{\mathcal{J}}_{y}(t)=\mathcal{J}_{y}(t)\cos\vartheta(t)-\mathcal{J}_{x}(t)\sin\vartheta(t),
 \end{eqnarray}
where
\begin{eqnarray}\label{eqA4}
\vartheta(t)=\int_{t_{0}}^{t}d\tau\Theta_{z}(\tau),
 \end{eqnarray}
is the phase accumulated by the system during a sweep. As a consequence, noise fields remains centered $\langle\tilde{\mathcal{J}}_{x,y}(t)\rangle=0$ but multiplicative: the transverse components of the new noise fields are statistically dependent. Equations (\ref{eqA2}) and (\ref{eqA3}) show that the gauge which eliminates fast oscillating terms rotates the coordinates frame of the transverse plane of an angle $\vartheta(t)$.

We construct a FP equation for the probability distribution function\cite{Risken} 
\begin{eqnarray}\label{eqA7}
\mathbf{W}(Q,t)=\langle\delta(Q-Q(t))\rangle,
\end{eqnarray}
which satisfies the initial condition $\mathbf{W}(Q,t_{0})=\delta(Q-Q_{0})$.
The brackets $\langle...\rangle$ denote an averaging over all possible realizations of the noise. If one writes the equation for the density $\delta(Q-Q(t))$ of the system in the configuration $Q(t)$ at time $t$ and average the resulting equation over the noise realizations\cite{Risken}:
\begin{eqnarray}\label{eqA8}
\frac{\partial \mathbf{W}}{\partial t}+\frac{\partial}{\partial Q_{\mu}} A_{\mu}(Q,t)\mathbf{W}=-\frac{\partial}{\partial Q_{\mu}}B_{\mu k}(Q,t)
\langle\tilde{\mathcal{J}}_{k}(t)\delta(Q-Q(t))\rangle.
\end{eqnarray}
The remaining average in this equation is simplified with the help of Novikov's theorem\cite{Novikov} and after some simplifications yields
\begin{eqnarray}\label{eqA9}
\langle\tilde{\mathcal{J}}_{k}(t)\delta(Q-Q(t))\rangle=-\frac{\partial}{\partial Q_{\nu}}\langle\mathcal{J}_{k\nu}(t) \delta(Q-Q(t))\rangle,
\end{eqnarray}
where
\begin{eqnarray}\label{eqA10}
\mathcal{J}_{k\nu}(t)=\int_{t_{0}}^{t}dt_{1}\mathcal{G}_{k m}(\gamma|t-t_{1}|) 
 \frac{\delta Q_{\nu}(t)}{\delta \tilde{\mathcal{J}}_{m}(t_{1})},
\end{eqnarray}
and $\mathcal{G}_{k m}(\gamma|t-t_{1}|)$ are two-time correlation functions for the noise fields $\tilde{\mathcal{J}}_{\nu}$ that can be derived from (\ref{eqA2}) and (\ref{eqA3}) considering (\ref{equ39}). Due to the random character of noise in the transverse direction, $\tilde{\mathcal{J}}_{z}(t)=0$ and $\mathcal{J}_{3\nu}(t)$=0.  The relation (\ref{eqA10}) is valid for arbitrary zero-mean noise field with Gaussian realizations.

When the noise is fast, its characteristic correlation time $1/\gamma$ is small compared to the transition 
 time of the system. We consider the noise contribution at the vicinity of $t_{1}\approx t$. The Taylor series expansion of $\delta Q_{\nu}(t)/\delta \tilde{\mathcal{J}}_{m}(t_{1})$ is considered,
 \begin{eqnarray}\label{eqA11}
  \frac{\delta Q_{\nu}(t)}{\delta \tilde{\mathcal{J}}_{m}(t_{1})}\approx\frac{\delta Q_{\nu}(t)}{\delta \tilde{\mathcal{J}}_{m}(t_{1})}\Big|_{t_{1}=t}+
  \frac{d}{dt_{1}}\frac{\delta Q_{\nu}(t)}{\delta \tilde{\mathcal{J}}_{m}(t_{1})}\Big|_{t_{1}=t}\tau+\mathcal{O}[\tau^2].
\end{eqnarray}
Here, $\tau=t_{1}-t$ is a characteristic time interval in which the coherence of the system is preserved. On the other hand $\tau$ is  small enough as  $\gamma\to\infty$. This permits to consider only the first term in the expansion (\ref{eqA11}).  We readily integrate the Langevin equation  to 
 find this term and we obtain
\begin{eqnarray}\label{eqA12}
Q_{\nu}(t)=\int_{t_{0}}^{t} dt' A_{\nu}(Q',t')+\int_{t_{0}}^{t}dt'B_{\nu k}(Q',t')\tilde{\mathcal{J}}_{k}(t'),\hspace{0.5cm}\quad
\end{eqnarray}
since  $Q_{\nu}(t_{0})=0$. The functional derivative of equation (\ref{eqA12}) yields:
 \begin{eqnarray}\label{eqA13}
\nonumber\frac{\delta Q_{\nu}(t)}{\delta \tilde{\mathcal{J}}_{m}(t_{1})}=\int_{t_{1}}^{t} dt'\frac{\delta A_{\nu}(Q',t')}{\delta \tilde{\mathcal{J}}_{m}(t_{1})}
  +\int_{t_{1}}^{t} dt'\frac{\delta B_{\nu k}(Q',t')}{\delta\tilde{\mathcal{J}}_{m}(t_{1})}\tilde{\mathcal{J}}_{k}(t')+B_{\nu m}(Q(t_{1}),t_{1}),\\
\end{eqnarray}
where 
 \begin{eqnarray}\label{eqA14}
\frac{\delta Q_{\nu}(t)}{\delta \tilde{\mathcal{J}}_{m}(t_{1})}\approx\frac{\delta Q_{\nu}(t)}{\delta \tilde{\mathcal{J}}_{m}(t_{1})}\Big|_{t=t_{1}}=B_{\nu m}(Q(t),t).
\end{eqnarray}
The integral in (\ref{eqA10}) can be evaluated. To do this, we first calculate all the two-time correlation functions $\mathcal{G}_{km}(...)$ in equation (\ref{eqA10}). From equation (\ref{equ39}), we write the function $\mathcal{R}(\gamma|t-t'|)=\eta^{2}\exp(-\gamma|t-t'|)$ that permits the calculation of the following
\begin{eqnarray}\label{eqA15a}
\mathcal{J}_{1\nu}(t)\approx \nonumber B_{\nu 1}(Q(t),t)\int_{0}^{\infty}d\tau\mathcal{R}(\gamma|\tau|)\cos[\vartheta(t)-\vartheta(t+\tau)]+\\\hspace{3cm} B_{\nu 2}(Q(t),t)\int_{0}^{\infty}d\tau\mathcal{R}(\gamma|\tau|)\sin[\vartheta(t)-\vartheta(t+\tau)],
\end{eqnarray}
and
\begin{eqnarray}\label{eqA15b}
\mathcal{J}_{2\nu}(t)\approx \nonumber -B_{\nu 1}(Q(t),t)\int_{0}^{\infty}d\tau\mathcal{R}(\gamma|\tau|)\sin[\vartheta(t)-\vartheta(t+\tau)]+\\\hspace{3cm} B_{\nu 2}(Q(t),t)\int_{0}^{\infty}d\tau\mathcal{R}(\gamma|\tau|)\cos[\vartheta(t)-\vartheta(t+\tau)],
\end{eqnarray}

As $\tau$ is small, the phase $\vartheta(t+\tau)$ is expanded as a series of $\tau$ up to the linear term $\vartheta(t+\tau)\approx\vartheta(t)+\frac{\partial\vartheta(t)}{\partial t}\tau+\mathcal{O}(\tau^{2})$. Therefore, we deduce $\vartheta(t)-\vartheta(t+\tau)\approx-\omega(t)\tau$ where the frequency $\omega(t)$ reads
\begin{eqnarray}\label{eqA16}
\omega(t)=\frac{\partial\vartheta (t)}{\partial t}=\Theta_{z}(t).
\end{eqnarray}
From equations (\ref{eqA15a}) and (\ref{eqA15b}) we have
\begin{eqnarray}\label{eqA17}
\mathcal{J}_{k\nu}(t)\approx B_{\nu k}(Q,t)\Omega(t),
\end{eqnarray}
where
\begin{eqnarray}\label{eqA18}
\Omega(t)=\int_{0}^{\infty}d\tau\mathcal{R}(\gamma|\tau|)\cos[\omega(t)\tau].
\end{eqnarray}
 Finally, a FP equation associated with the noise is written as: 
\begin{eqnarray}\label{eqA19}
\frac{\partial\mathbf{W}}{\partial t}=-\frac{\partial}{\partial Q_{\mu}}D_{\mu}(Q,t)\mathbf{W}+
\frac{\partial^{2}}{\partial Q_{\mu}\partial Q_{\nu}}D_{\mu\nu}(Q,t)\mathbf{W},
\end{eqnarray}
where the drift and diffusion coefficients are respectively given by
\begin{eqnarray}\label{eqA20}
D_{\mu}(Q,t)=A_{\mu}(Q,t)+\frac{\partial B_{\mu k}(Q,t)}{\partial Q_{\nu}} \mathcal{J}_{k\nu}(t) ,
\end{eqnarray}
and 
\begin{eqnarray}\label{eqA21}
D_{\mu\nu}(Q,t)=B_{\mu k}(Q,t)\mathcal{J}_{k\nu}(t).
\end{eqnarray}
For our case, the procedure for elimination of fast oscillating terms in equation (\ref{equ5}) are indicated in equation (\ref{equ40}). $Q_{\mu}(t)=[\tilde{f}(t),\tilde{h}(t),\tilde{g}(t)]$, $\mathcal{J}_{1,2}(t)=\mathcal{J}_{x,y}(t)$ and $\mathcal{J}_{z}(t)=0$.  The column vector $A_{\mu}(Q,t)$ vanishes while the elements $B_{\mu k}(Q,t)$ form a ($3\times3$) matrix.
$\Theta_{z}(t)=\omega(t)=2vt$.  From equation (\ref{eqA19}), after a long calculation, we derive  the FP equation in  (\ref{equ43}).  

\section*{Acknowledgments}
This work is supported by the Sandwich Training Educational Programme (STEP) of the Abdus Salam International 
Centre for Theoretical Physics (ICTP), Trieste, Italy.

\end{document}